\documentclass[12pt,a4paper,aps,preprint,superscriptaddress,nofootinbib]{revtex4-1}
\usepackage{graphicx}
\usepackage{hyperref}
\usepackage{cancel}
\usepackage{amssymb}
\usepackage{textcomp}
\usepackage{amsmath}
\usepackage{cases}
\usepackage{bm}
\usepackage{times}
\usepackage{epsfig}
\usepackage{color}
\usepackage{mathrsfs}
\usepackage{esint}
\usepackage{ulem}
\usepackage{subfigure}
\usepackage{makecell}
\usepackage{booktabs}
\makeatother

\newcommand{\calG}{{\cal G}}

\begin{document}
\title{\LARGE On the anisotropies of the cosmological gravitational-wave background from pulsar timing array observations}
\bigskip
\author{Ran Ding}
\email{dingran@mail.nankai.edu.cn}
\author{Chi Tian}
\email{ ctian@ahu.edu.cn  (Corresponding author).}
\affiliation{
School of Physics and Optoelectronics Engineering, Anhui University, 111 Jiulong Road, Hefei, Anhui, China 230601}
\date{\today}

\begin{abstract}
Significant evidence for a stochastic gravitational-wave background has recently been reported by several Pulsar Timing Array observations. These studies have shown that, in addition to astrophysical explanations based on supermassive black hole binaries (SMBHBs), cosmological origins are considered equally important sources for these signals.
To further explore these cosmological sources, in this study, we discuss the anisotropies in the cosmological gravitational wave background (CGWB) in a model-independent way. Taking the North American Nanohertz Observatory for Gravitational Waves (NANOGrav) 15-year dataset as a benchmark, we estimate the angular power spectra of the CGWB and their cross-correlations with cosmic microwave background (CMB) fluctuations and weak gravitational lensing. We find that the NANOGrav 15-year data implies suppressed Sachs-Wolf (SW) effects in the CGBW spectrum, leading to a marginally negative cross-correlation with the CMB at large scales.  This procedure is applicable to signals introduced by different early universe processes and is potentially useful for identifying unique features about anisotropies of CGWB from future space-based interferometers and astrometric measurements.
\end{abstract}

\maketitle

\section{Introduction}

The stochastic gravitational wave background (SGWB) is the diffuse gravitational wave (GW) signal produced by the superposition of numerous unresolved sources~\cite{Romano:2016dpx,Christensen:2018iqi,Renzini:2022alw,vanRemortel:2022fkb,LISACosmologyWorkingGroup:2022jok}. Based on its different origins, the SGWB is generally divided into two categories: the cosmological GW background  (CGWB) and the astrophysical GW background (AGWB). The CGWB originated from the early Universe processes are closely related to new physics, such as inflation, reheating/preheating after inflation, early primordial black holes,  topological defects, and first-order phase transitions~\cite{Vachaspati:1984gt,Kosowsky:1992rz,Guzzetti:2016mkm,Caprini:2018mtu,Aggarwal:2020olq}. On the other hand, the AGWB is generated in the matter dominated era through activities of astronomical objects, including massive black hole binaries, extreme/intermediate mass ratio inspirals and compact binaries in the Milk Way~\cite{Ferrari:1998jf,Ferrari:1998ut,Schneider:2000sg,Farmer:2003pa,Regimbau:2011rp}. The rich information carried by the SGWB makes it a powerful probe of both the early Universe and large scale structure of matter distribution. As a result, it has become one of the primary targets of present and future gravitational wave detections.

Notably, recent observations of the SGWB by pulsar timing arrays (PTA) from NANOGrav~\cite{NANOGrav:2023gor,NANOGrav:2023hde}, CPTA~\cite{Xu:2023wog}, EPTA~\cite{EPTA:2023sfo,EPTA:2023akd,EPTA:2023fyk}
, and PPTA~\cite{Reardon:2023gzh,Reardon:2023zen,Zic:2023gta} have provided substantial evidence for the SGWB. These studies have also performed comprehensive analysis on various typical CGWB and AGWB interpretations to the detected signal~\cite{NANOGrav:2023hvm,NANOGrav:2023hfp,EPTA:2023xxk}. Despite distinct different frequency dependencies of the CGWB and AGWB, these studies have not identified significant evidence to support either of them as the exclusive source of the SGWB. In addition, ongoing ground-based interferometers, such as the aLIGO/Virgo/KAGRA collaboration, are close to reaching the sensitivity required to detect the AGWB~\cite{LIGOScientific:2016jlg,KAGRA:2021kbb}. Future space-based telescopes, such as LISA~\cite{LISA:2017pwj, LISA:2022kgy}, Taiji~\cite{Hu:2017mde,Ruan:2018tsw}, TianQin~\cite{TianQin:2015yph}, BBO~\cite{Corbin:2005ny} and DECIGO~\cite{Kawamura:2006up}, as well as ground-based interferometers, such as Einstein Telescope~\cite{Maggiore:2019uih} and Cosmic Explorer~\cite{Reitze:2019iox}, have the potential ability to detect the SGWB generated by various mechanisms~\cite{Kuroyanagi:2018csn,Caprini:2019pxz}. 

The SGWB also exhibits anisotropies, and these anisotropies in the SGWB should have noticeable discrepancies in their spectra between the CGWB~\cite{Alba:2015cms,Contaldi:2016koz,Bartolo:2019oiq,Bartolo:2019yeu,ValbusaDallArmi:2020ifo,Li:2021iva,Schulze:2023ich,Cui:2023dlo,Wang:2023ost, Li:2023qua, ValbusaDallArmi:2023ydl, Bethke:2013aba} and AGWB~\cite{Cusin:2017fwz,Cusin:2018rsq,Jenkins:2018uac,Jenkins:2018kxc,Cusin:2019jpv,Bertacca:2019fnt,Bellomo:2021mer,Sato-Polito:2023spo, Pol:2022sjn} due to their different sourcing and propagating mechanisms. For instance, compared to the AGWB, the CGWB generally began propagating at earlier times, involving more deeply with various cosmological perturbations during its propagation, leading to a distinct angular spectrum of anisotropies. Therefore, with the enhanced angular resolutions of these future GW detectors, the anisotropies in the SGWB could potentially serve as an independent probe for distinguishing GW sources. 
The production and propagation of the CGWB in the perturbed universe shares the similar physics with the CMB photons, making these two signals correlated. This correlation serves as another important observable and can be useful in multiple ways, such as testing primordial \cite{Li:2023qua} or isocurvature perturbations \cite{Geller:2018mwu}, examining CMB anomalies\cite{Galloni:2022rgg}, or improving the signal-to-noise ratio \cite{Alonso:2020mva}.
Therefore, it is worth estimating the anistropic information based on existing PTA observations and examining the level of cross-correlations with other cosmological tracers.

In this study, we focus on the CGWB origin, taking the NANOGrav 15-year data set~\cite{NANOGrav:2023gor} as a benchmark, and discuss its implications on the anisotropies of SGWB in a model-independent way.
This paper is structured as follows. In Sec.~\ref{sec:ani}, we perform a Bayesian analysis on the recent NANOGrav 15-year data and outline the procedure for modeling the anisotropies of the CGWB. We then calculate the specific CGWB angular power spectrum based on the NANOGrav data. In Sec.~\ref{sec:cross}, we further discuss on the cross-correlations between the CGWB and other CMB tracers. We summarize and conclude in Sec.~\ref{sec:diss}.
Throughout this paper, we assume that the standard $\Lambda \mathrm{CDM}$ cosmology with flat spatial curvature. Corresponding cosmological parameter values are adopt from the Planck 2018 results~\cite{Planck:2018vyg}.

\section{CGWB anisotropies based on PTA observations}
\label{sec:ani}
\begin{figure}
\centering
\includegraphics[width=0.9\textwidth]{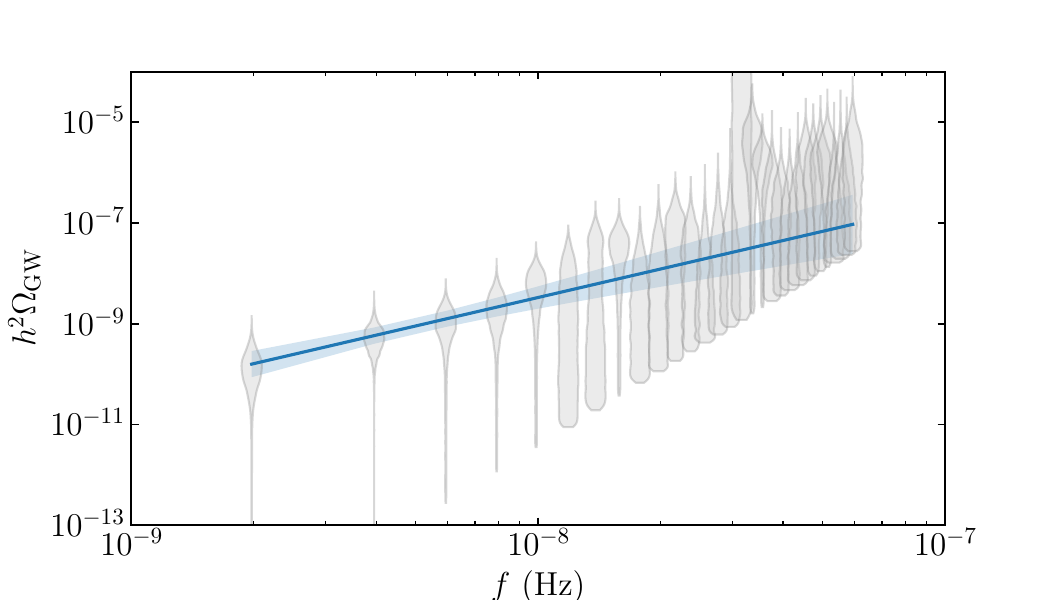}
\includegraphics[width=0.6\textwidth]{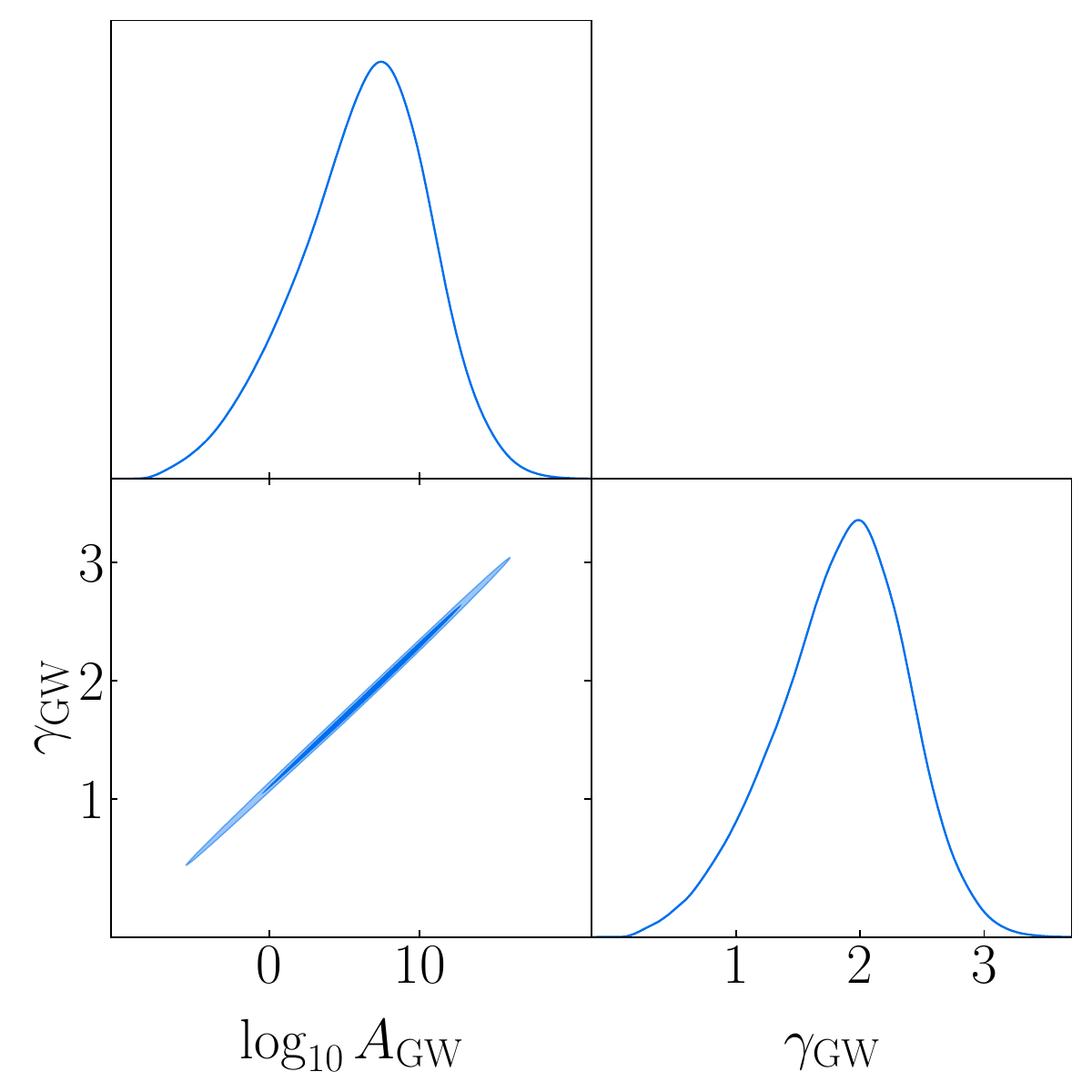}
\caption{\textbf{Upper panel}: the best-fit power-law GWB spectrum (blue solid line) with the results of the free spectrum with the Hellings-Down correlation from NANOGrav 15-year data set (gray violins). The light blue band is computed based on $\gamma_{\rm{GW}}$ ranging within $68\%$ confidence interval. \textbf{Lower panel}: triangle plots of posteriors.}
\label{fig:spectrum}
\end{figure}

Current PTA observations reveal the information of GWB spectrum at $f \sim \mathcal{O}(\mathrm{nHz})$ frequency bands, suggesting a near power-law relation between the GW density fraction and the frequencies. This is consistent with most of the CGWB models~\cite{Kuroyanagi:2018csn,NANOGrav:2023gor,NANOGrav:2023hvm}. It is therefore reasonable to take a model-independent approach. In this study, we consider the NANOGrav 15-year data set that consists of measurements of 68 millisecond pulsars measured between 2004 July and 2020 August~\cite{NANOGrav:2023hde}. We parameterize the GWB spectrum as a power-law function $\Omega_{\rm{GW}}h^2=A_{\rm{GW}}(f/\rm{Hz})^{\gamma_{\rm{GW}}}$, and take the amplitude $A_{\rm{GW}}$ and slope as $\gamma_{\rm{GW}}$ free parameters. Using the results of the free spectrum from the Hellings-Down correlation, we perform the Bayesian fitting with the help of Markov chain Monte Carlo (MCMC) sampler ${\tt emcee}$~\cite{Foreman_Mackey_2013}.  Based on lowest 14 frequency bins, the best-fit values of amplitude and slope with $68\%$ confidence regions are given by
\begin{align}
\log_{10}A_{\rm{GW}}=6.02_{-3.34}^{+2.62}, \quad \gamma_{\rm{GW}}=1.82_{-0.40}^{+0.32},
\label{eq:1}
\end{align}
corresponding GWB spectrum and triangle plots of $A_{\rm{GW}}$ and $\gamma_{\rm{GW}}$ shown in Fig.~\ref{fig:spectrum}.

Equipped with this model independent GWB spectrum that fits the NANOGrav 15-year data, we now discuss its implications of anisotropies in CGWB. The GW content is described by a distribution function $f_{\rm{GW}}(\vec{x}, p, \hat{p}, \eta)$, as a function of position $\vec{x}$, conformal time $\eta$, comving momentum/frequency $p=aE$ and direction of propagation $\hat{p}^i=p^i/p$. Obviously, $f_{\rm{GW}}$ is specified by the mechanism that generates the GW, and can have either thermal or non-thermal distributions. The total energy density of GW is obtained by integrating the distribution function over the three-momentum
\begin{align}
\rho_{\rm{GW}}(\vec{x}, \eta) &=\int d^3\vec{p} \,p f_{\rm{GW}}(\vec{x}, p, \hat{p}, \eta) = \int dp d\Omega_{\hat{p}} \, p^3 f_{\rm{GW}}(\vec{x}, p, \hat{p}, \eta) \nonumber\\
&=\int d\ln p \, d\Omega_{\hat{p}} \, p^4 f_{\rm{GW}}(\vec{x}, p, \hat{p}, \eta),
\label{eq:2}
\end{align}
where $d\Omega_{\hat{p}}$ is the solid angular element in momentum space. The GW spectrum, as its definition, is expressed as
\begin{align}
\Omega_{\rm{GW}}(\vec{x}, p, \eta) \equiv \frac{1}{\rho_{\rm{c}}}\frac{d \rho_{\rm{GW}}}{d\ln p} = \int d\Omega_{\hat{p}} \frac{p^4}{\rho_{\rm{c}}}  f_{\rm{GW}}(\vec{x}, p, \hat{p}, \eta),
\label{eq:3}
\end{align}
where $\rho_{\rm{c}}=3 H^2/8 \pi G$ is the critical density and $H$ the Hubble rate. Following conventions in calculations of CMB anisotropy, we expand $f_{\rm{GW}}$ with respect to momentum $p$ to the first order~\footnote{The minus sign is included here to conform with the convention used for CMB anisotropies. By substituting the definition  into the Planck distribution of photon, given by $f \propto\left[\exp \left(\frac{p}{a T(\eta)}\right)-1\right]^{-1}$, and expanding to first order, we obtain CMB the temperature
perturbation $\Theta (x,\vec{x}, p, \hat{p}, \eta)=\delta{T}/T$.}~\cite{Bartolo:2019oiq},
\begin{align}
f_{\rm{GW}}(\vec{x}, p, \hat{p}, \eta)=\bar{f}_{\rm{GW}}(p) - p\frac{\partial \bar{f}_{\rm{GW}}}{\partial p} \calG (\vec{x}, p, \hat{p}, \eta),
\label{eq:4}
\end{align}
where $\bar{f}_{\rm{GW}}$ is the background distribution function which does not explicitly depend on time, and $\calG$ the linear perturbation of the distribution function. We also separate the total spectrum into the background part $\overline{\Omega}_{\mathrm{GW}}$ and the linear fluctuation  $\delta_{\rm{GW}}$:
\begin{align}
\Omega_{\rm{GW}}(\vec{x}, p, \eta) &= \int d\Omega_{\hat{p}} \frac{p^4}{\rho_{\rm{c}}}\bar{f}_{\rm{GW}} - \int d\Omega_{\hat{p}} \frac{p^5}{\rho_{\rm{c}}}\frac{\partial \bar{f}_{\rm{GW}}}{\partial p} \calG (\vec{x}, p, \hat{p}, \eta) \nonumber\\
&\equiv \overline{\Omega}_{\mathrm{GW}}(p, \eta)\left[1 + \delta_{\rm{GW}}(\vec{x}, p, \eta)\right].
\label{eq:5}
\end{align}
Then, by combining Eqs.~\eqref{eq:3}\eqref{eq:4}\eqref{eq:5},it is easy to prove the following relation
\begin{align}
\frac{\partial \ln \overline{\Omega}_{\mathrm{GW}}\left(p, \eta\right)}{\partial \ln p} \calG (\vec{x}, p, \hat{p}, \eta) &= \frac{1}{\overline{\Omega}_{\rm{GW}}}\left[4\int d\Omega \frac{p^4}{\rho_{\rm{c}}} \bar{f}_{\rm{GW}}\calG + \int d\Omega \frac{p^5}{\rho_{\rm{c}}}\frac{\partial \bar{f}_{\rm{GW}}}{\partial p}\calG\right] \nonumber\\
&= \frac{1}{\overline{\Omega}_{\rm{GW}}}\left[ 4 \overline{\Omega}_{\rm{GW}}\calG - \overline{\Omega}_{\rm{GW}} \delta_{\rm{GW}}\right] \nonumber\\
&= 4 \calG(\vec{x}, p, \hat{p}, \eta) - \delta_{\rm{GW}}(\vec{x}, p, \hat{p}, \eta),
\label{eq:6}
\end{align}
which allows us to express the fluctuation $\delta_{\rm{GW}}$ in terms of background spectrum and the linear perturbation of the distribution function~\cite{Bartolo:2019oiq,Bartolo:2019yeu}
\begin{align}
\delta_{\rm{GW}}(\vec{x}, p, \hat{p}, \eta) = \left[4-\frac{\partial \ln \overline{\Omega}_{\mathrm{GW}}\left(p, \eta\right)}{\partial \ln p}\right] \calG (\vec{x}, p, \hat{p}, \eta).
\label{eq:7}
\end{align}
Eq.~(\ref{eq:7}) establishes a direct connection between the anisotropies of GWB and the monopole component, where the later can be extracted from observations, making it useful for model-independent analysis. In  particular, the power-law spectrum considered in this work implies $\partial \ln \overline{\Omega}_{\rm{GW}}\left(p, \eta\right)/\partial \ln p = \gamma_{\rm{GW}}$, which is independent of the GW momentum and the conformal time.

In analogous to the CMB anisotropies, we can expand the perturbation function $\calG (\vec{x}, p, \hat{p}, \eta)$ with spherical harmonic basis: 
\begin{align}
\calG (\vec{x}, p, \hat{p}, \eta) &\equiv \sum_{\ell=1}^{\infty} \sum_{m=-\ell}^{\ell} a^{\rm GW}_{\ell m}(\vec{x}, p, \eta) Y_{\ell m}(\hat{p}), \label{eq:8}\\
a^{\rm GW}_{\ell m}(\vec{x}, p, \eta) &= \int d\Omega_{\hat{p}} Y_{\ell m}^*(\hat{p})\calG(\vec{x}, p, \hat{p},\eta). \label{eq:9}
\end{align}
 It is also useful to expand $\calG$ into multipole moments $\calG_l$ in Fourier space:
\begin{align}
\calG (\vec{k}, p, \mu, \eta) &\equiv \sum_{\ell=0}^{\infty}(-i)^{\ell}(2 \ell + 1) \calG_\ell (\vec{k}, p, \eta) \mathcal{P}_{\ell}(\mu), \label{eq:10}\\
\calG_\ell(\vec{k}, p, \eta) &= \frac{i^l}{2} \int_{-1}^1 d \mu \mathcal{P}_{\ell}(\mu) \calG (\vec{k}, p, \mu, \eta ), \label{eq:11}
\end{align}
where $\mathcal{P}_{\ell}(\mu)$ is the Legendre polynomials, and $\mu \equiv \hat{k} \cdot \hat{p} $ is the relative orientation between $\vec{k}$ and $\vec{p}$.
The auto-correlation function $\Bigl\langle a^{\rm GW}_{\ell m} a^{\rm GW *}_{\ell' m'}\Bigr\rangle= \delta_{\ell\ell'}\delta_{mm'}C^{\rm GW}_\ell$ gives the angular power spectrum, where the brackets denotes an ensemble average. We here only consider the Gaussian fluctuations, then the expansion coefficients $C^{\rm GW}_\ell$ contains all the information of CGWB anisotropies. In the linear perturbation theory, it is conventional to define a transfer function to connect $\calG_\ell(\vec{k}, \eta_0)$ and initial curvature perturbation $\mathcal{R}(\vec{k})$ as
\begin{align}
\calG_\ell(\vec{k}, p, \eta_0) = T^{\rm GW}_\ell(\vec{k}, p, \eta_{\rm{in}}, \eta_0)\mathcal{R}(\vec{k}),
\label{eq:12}
\end{align}
where $\eta_{\rm{in}}$ and $\eta_0$ respectively denote the conformal time at initial and the present time. The transfer function, generally depending on perturbations in matter, radiation and the metric at $\eta_{\rm{in}}$, describes the evolution of the primordial perturbations $\mathcal{R}(\vec{k})$. Using Eqs.~(\ref{eq:9}),~(\ref{eq:10}) and~(\ref{eq:12}), we can write down the ensemble average of harmonic coefficients as
\begin{align}
\Bigl \langle  a^{\rm GW}_{\ell m}a^{\rm GW *}_{\ell' m'} \Bigr \rangle &= \sum_{\ell_1,\ell_2} (2\ell_1 + 1)(2\ell_2+1)(-i)^{\ell_1-\ell_2}\int\frac{d^3kd^3k'}{(2\pi)^3}e^{i(\vec{k} - \vec{k}')\cdot\vec{x}} \Bigl \langle\calG_{\ell_1}(\vec{k}, p, \eta_0) \calG_{\ell_2}^*(\vec{k}', p, \eta_0) \Bigr \rangle \nonumber\\
&\times \int d\Omega_{\hat{p}} \mathcal{P}_{\ell_1}(\mu) Y_{\ell m}^*(\hat{p}) \int d\Omega_{\hat{p}'} \mathcal{P}_{\ell_2}(\mu') Y_{\ell'm'}(\hat{p}')  \nonumber\\
&= \sum_{\ell_1,\ell_2} (2\ell_1 + 1)(2\ell_2+1)(-i)^{\ell_1-\ell_2}\int\frac{d^3k}{(2\pi)^3}\frac{2\pi^2}{k^3}\mathcal{P}_\mathcal{R}(k) T^{\rm GW}_{\ell_1}(k, p, \eta_{\rm{in}}, \eta_0) T^{\rm GW}_{\ell_2}(k, p, \eta_{\rm{in}}, \eta_0) \nonumber\\
&\times \int d\Omega_{\hat{p}} \mathcal{P}_{\ell_1}(\mu) Y_{\ell m}^*(\hat{p}) \int d\Omega_{\hat{p}'} \mathcal{P}_{\ell_2}(\mu') Y_{\ell'm'}(\hat{p}')  \nonumber\\
&= \delta_{\ell\ell'}(4\pi)^2\int\frac{d^3k}{(2\pi)^3} \frac{2\pi^2}{k^3}\mathcal{P}_\mathcal{R}(k) Y_{\ell m}(\hat{k})Y_{\ell m'}^*(\hat{k}) \Big|T^{\rm GW}_\ell(k, \eta_{\rm{in}}, \eta_0) \Big|^2  \nonumber\\
&= \delta_{\ell\ell'}\delta_{mm'}4\pi\int\frac{dk}{k}\mathcal{P}_\mathcal{R}(k) \Big|T^{\rm GW}_\ell(k, p, \eta_{\rm{in}}, \eta_0)\Big|^2,
\label{eq:13}
\end{align}
where we have used the definition of the primordial power-spectrum
\begin{align}
\Bigl \langle\mathcal{R}(\vec{k})\mathcal{R}^*(\vec{k}') \Bigr \rangle = \delta(\vec{k} - \vec{k}')\frac{2\pi^2}{k^3}\mathcal{P}_\mathcal{R}(k), 
\end{align}
in the second line, and the relations
\begin{align}
\int d\Omega_{\hat{p}} \mathcal{P}_\ell(\mu) Y_{\ell'm}^*(\hat{p}) = \delta_{\ell\ell'} \frac{4\pi}{2\ell+1} Y^*_{\ell m}(\hat{k}), \quad \int d\Omega_{\hat{k}} Y_{\ell m}(\hat{k})Y_{\ell m'}^*(\hat{k}) = \delta_{mm'},
\end{align}
in the third and last line, respectively.

From Eq.~(\ref{eq:13}), following the definition, the angular power spectrum of the CGWB anisotropies is
\begin{align}
    C^{\rm GW}_\ell = 4\pi\int\frac{dk}{k}\mathcal{P}_\mathcal{R}(k) \Big|T^{\rm GW}_\ell(k, p, \eta_{\rm{in}}, \eta_0) \Big|^2.
\end{align}
Once a primordial power-spectrum $\mathcal{P}_\mathcal{R}(k)$ is given, the angular power spectrum is then entirely determined by the transfer function $T^{\rm GW}_\ell(k, p, \eta_{\rm{in}}, \eta_0)$, which can be calculated by solving the Boltzmann equation of GWs in the perturbed Friedmann-Robertson-Walker (FRW) background. As shown in the pioneering works~\cite{Isaacson:1968hbi,Isaacson:1968zza,Misner:1973prb}, one can take geometric optics approximation when GWs fulfill the shortwave limit, and its propagation is described by Boltzmann equation~\cite{Alba:2015cms,Contaldi:2016koz,Bartolo:2019oiq,Bartolo:2019yeu},
\begin{align}
\frac{d f_{\rm GW}}{d \eta}=\mathcal{C}[f_{\rm GW}(x^\mu, p^\mu)]+\mathcal{J}[f_{\rm GW}(x^\mu, p^\mu)],
\label{eq:16}
\end{align}
where $\mathcal{C}$ and $\mathcal{J}$  stand for the collision term and the emission term respectively. The collision term can be safely neglected since GWs are decoupled from other components of the Universe. As for the emission term, which accounts for the emissitivity from cosmological/astrophysical sources, can be treated as an initial condition for the GW distribution in the case of a CGWB. As a consequence, similar to photons, GWs propagate along null geodesics, and Boltzmann equation~(\ref{eq:16}) simplifies to $d f_{\rm GW}/d \eta=0$. In the conformal Newtonian gauge, the FRW metric with first order scalar perturbations is written as
\begin{align}
d s^2=a^2(\eta)\left[-( 1+ 2 \Psi) d \eta^2 + (1-2 \Phi) \delta_{i j} d x^i d x^j\right],
\end{align}
where $a(\eta)$ is the scale factor as a function of conformal time, $\Psi (\vec{x}, \eta)$ and $\Phi (\vec{x}, \eta)$ denote
scalar potential and curvature perturbations. Under this metric, the Boltzmann equation becomes~\cite{Bartolo:2019oiq,Bartolo:2019yeu}
\begin{align}
\frac{d f_{\mathrm{GW}}}{d \eta} = \frac{\partial f_{\mathrm{GW}}}{\partial \eta}+ \hat{p}^i \partial_i f_{\mathrm{GW}} - p \frac{\partial f_{\mathrm{GW}}}{\partial p}\left[\Phi^{\prime} -\hat{p}^i \partial_i \Psi\right]=0.
\label{eq:18}
\end{align}
In above equation, a prime denotes derivative with respect to the conformal time.  Inserting Eq.~(\ref{eq:4}) into Eq.~(\ref{eq:18}) and using the background Boltzmann equation $\partial \bar{f}_{\mathrm{GW}}/\partial \eta=0$, we derive the first order Boltzmann equation, expressed in Fourier space as
\begin{align}
\calG^{\prime}(k, p, \mu, \eta) +i k \mu \calG (k, p, \mu, \eta) = \Phi^{\prime}(k, \eta)-i k \mu \Psi (k, \eta).
\label{eq:19}
\end{align}
The physical meaning of the above equation is simple: the left-hand side and right-hand side terms respectively corresponding to the free-streaming and the gravitational effect. We can express the solution of Eq.~(\ref{eq:19}) by using the line-of-sight integration method~\cite{Seljak:1996is}. To this end, we first rewrite the free-streaming term as 
\begin{align}
\calG^{\prime}+i k \mu \calG=e^{-i k \mu (\eta - \eta_0)} \frac{d}{d \eta}\left[\calG e^{i k \mu (\eta - \eta_0)}\right].
\end{align}
Multiplying both sides by a factor $e^{i k \mu (\eta - \eta_0)}$ and integrate $\eta$ over $[\eta_{\rm{in}}, \eta_0]$, we arrive
\begin{align}
\calG\left(k, p, \mu, \eta_0\right)=\calG\left(k, p, \mu, \eta_{\rm{in}}\right) e^{i k \mu\left(\eta_{\rm {in}}-\eta_0\right)} + \int_{\eta_{\rm{in}}}^{\eta_0} d\eta \left[\Phi^{\prime} (k, \eta)-i k \mu \Psi (k, \eta)\right] e^{i k \mu\left(\eta-\eta_0\right)}.
\label{eq:21}
\end{align}
The last term on right-hand side of Eq.~\eqref{eq:21} can be recast into
\begin{align}
i k \mu \int_{\eta_{\rm{in}}}^{\eta_0} d\eta \Psi e^{i k \mu\left(\eta-\eta_0\right)} &= \int_{\eta_{\rm{in}}}^{\eta_0} d\eta \Psi \frac{d}{d \eta}\left[e^{i k \mu (\eta - \eta_0)}\right] \nonumber\\
&=\left.\Psi e^{i k \mu\left(\eta-\eta_0\right)}\right|_{\eta_{\rm{in}}}^{\eta_0}-\int_{\eta_{\rm{in}}}^{\eta_0} \Psi^{\prime} e^{i k \mu\left(\eta-\eta_0\right)} \nonumber\\
&= - \Psi(k, \eta_{\rm{in}}) e^{i k \mu\left(\eta_{\rm{in}}-\eta_0\right)}- \int_{\eta_{\rm{in}}}^{\eta_0} \Psi^{\prime}(k, \eta) e^{i k \mu\left(\eta-\eta_0\right)},
\end{align}
where we have dropped the boundary $\Psi$-term at $\eta_0$, which has no observational effect. Then Eq.~(\ref{eq:21}) becomes
\begin{align}
\calG \left(k, p, \mu, \eta_0\right)=\left[\calG (k, p, \mu, \eta_{\rm{in}}) + \Psi(k, \eta_{\rm{in}})\right]e^{i k \mu\left(\eta_{\rm{in}}-\eta_0\right)} +\int_{\eta_{\rm{in}}}^{\eta_0} d \eta \left[\Psi^{\prime}(k, \eta)+\Phi^{\prime}(k, \eta)\right] e^{i k \mu\left(\eta-\eta_0\right)}.
\label{eq:23}
\end{align}
We can proceed to derive the multipole expression. By multiplying Eq.~(\ref{eq:11}) by a factor of $\frac{i^{\ell}}{2} \mathcal{P}_{\ell}(\mu)$ and then integrating $\mu$ over $[-1,1]$, we get  
\begin{align}
\calG_\ell\left(k, p, \eta_0\right) & \simeq \underbrace{\left[\calG_0\left(k, p, \eta_{\rm{in }}\right)+\Psi\left(k, \eta_{\rm{in }}\right)\right]}_{\rm{SW}} j_l\left[k\left(\eta_0-\eta_{\rm{in }}\right)\right] \nonumber\\
& + \underbrace{\int_{\eta_{\rm{in }}}^{\eta_0} d \eta \left[\Psi^{\prime}(k, \eta)+\Phi^{\prime}(k, \eta)\right]}_{\rm{ISW}} j_l\left[k\left(\eta_0-\eta)\right]\right.,
\label{eq:24}
\end{align}
where we have employed the definition of the multipoles in Eq.~(\ref{eq:11}) and the relation between spherical Bessel functions and Legendre polynomials:  
$j_{\ell}(-x)=\frac{i^{\ell}}{2} \int_{-1}^1 \mathcal{P}_{\ell}(\mu) e^{i x \mu} d \mu$.
We have also kept the monopole component in $\calG_\ell\left(k, p, \eta_{\rm{in }}\right)$ and dropped higher multipoles on right-hand side of Eq.~(\ref{eq:24}). This is due to the fact that for a very small initial time $\eta_{\rm{in}}$ and superhorizon Fourier mode $k$, $\calG_0$ is the only important term in multipoles. To see this, using recursion relation of Legendre polynomials $\mu \mathcal{P}_{\ell}(\mu)=\frac{\ell+1}{2 \ell+1} \mathcal{P}_{\ell+1}(\mu)+\frac{\ell}{2 \ell+1} \mathcal{P}_{\ell-1}(\mu)$, the free-streaming term $\calG^{\prime}+i k \mu \calG$ then implies that the multipoles follow the hierarchy $\calG_\ell \sim k\eta \calG_{\ell-1}$. For the initial condition we are interested in, i.e., $k\eta_{\rm{in}} \ll 1$, this hierarchical structure indicates that all higher-order moments with $\ell \ge 2$ are suppressed and can be safely neglected. This feature is in some way similar to the CMB anisotropies, in which photon multipoles with $\ell \ge 2$ are suppressed by photon optical depth in the tight-coupling regime. Moreover, it can be proven that $\calG_1\left(k, p, \eta_{\rm{in }}\right) \ll \Psi\left(k, \eta_{\rm{in }}\right)$ holds true for the adiabatic initial condition~\cite{Dodelson:2003ft}, making the dipole moment negligible as well. In the case of adiabatic perturbations, the initial condition of monopole component $\calG_0\left(k, \eta_{\rm{in }}\right)$ can be further connected to metric perturbation $\Psi\left(k, \eta_{\rm{in}}\right)$. To see this, consider the time-time component of background and linear perturbation Einstein equation during radiation-dominated era,
\begin{numcases}{}
3 \mathcal{H}^2 = 8 \pi G a^2 \bar{\rho} \label{eq:25}\\
k^2 \Phi-3 \mathcal{H}(\Phi^\prime+\mathcal{H} \Psi) = 4 \pi G a^2 \delta \rho, \label{eq:26}
\end{numcases}
where $\bar{\rho}\equiv \sum_i \rho_i$ and $\delta \rho \equiv \sum_i \delta \rho_i$ are respectively the total density and density perturbation, $i$ sums over all of relativistic components during radiation-dominated era, and $\mathcal{H}$ the Hubble rate as a function of conformal time. On super horizon scales $k \ll \mathcal{H}$, using the feature that $\Phi$ is constant ~\footnote{During the radiation-dominated era, the equation of curvature perturbation reads
$
\Phi^{\prime\prime} +\frac{4}{\eta} \Phi^\prime = \frac{1}{3} k^2\Phi.
$
On super-horizon scales, we can drop the term on the right-hand side, implying $\Phi =\rm const$.} and in adiabatic condition all relativistic species have same the fractional perturbations, by combining Eqs.~(\ref{eq:25}) and~(\ref{eq:26}), the initial density perturbation then reads $\delta_i \simeq -\frac{2}{\mathcal{H}}\left(\Phi^{\prime}+\mathcal{H} \Psi\right)= -2 \Psi$. From the relation between the GW density perturbation $\delta_{\rm{GW}}$ and $\calG$ in Eq.~(\ref{eq:7}), we have
\begin{align}
\calG_0\left(k, \eta_{\rm{in }}\right)\simeq -\frac{2 \Psi\left(k, \eta_{\rm{in}}\right)}{4-\gamma_{\rm{GW}}}.
\label{eq:27}
\end{align}
Using above relation to replace $\calG_0$ in Eq.~(\ref{eq:24}) and compare with Eq.~(\ref{eq:12}), the relevant transfer function can be extracted as
\begin{align}
T^{\rm GW}_\ell(k, \eta_{\rm{in}}, \eta_0) &= T_{\ell}^{\mathrm{SW}}\left(k, \eta_{\rm{in}}, \eta_0\right) + T_{\ell}^{\mathrm{ISW}}\left(k, \eta_{\rm{in}}, \eta_0\right), \label{eq:28}\\
T_{\ell}^{\mathrm{SW}}\left(k, \eta_{\rm{in}}, \eta_0\right) &= \left[1-\frac{2}{4-\gamma_{\rm{GW}}} \right]\frac{\Psi(k, \eta_{\rm{in}})}{\mathcal{R}(\vec{k})}j_{\ell}\left[k\left(\eta_0-\eta_{\text {in }}\right)\right], \label{eq:29}\\
T_{\ell}^{\mathrm{ISW}}\left(k, \eta_{\rm{in}}, \eta_0\right) &= \int_{\eta_{\rm{in}}}^{\eta_0} d \eta \frac{\left[\Psi^{\prime}(k, \eta) + \Phi^{\prime}(k, \eta)\right]}{\mathcal{R}(\vec{k})}j_{\ell}\left[k\left(\eta_0-\eta\right)\right], \label{eq:30}
\end{align}
and the GW angular power spectrum is obtained through Eq.~(\ref{eq:13}).

Before presenting our numerical results of the angular power spectrum, it is worth to make the following remarks:
\begin{itemize}
  \item {The GW transfer function shares some similar features with ones in CMB analysis, i.e., the propagation of GW is characterized by the summation of Sachs-Wolfe (SW) and Integrated Sachs-Wolfe (ISW) contributions. SW term represents the intrinsic anisotropy of GWs and its gravitational redshift when GWs escape from a potential well $\Psi$ at the surface $\eta_{\rm{in}}$; and ISW term accounts for the effect of gravitational redshift from the time evolution of the $\Psi+\Phi$ term along the line-of-sight. However, there are two notable differences:
\begin{enumerate}
\item {There is no Doppler term in the CGWB because GWs are decoupled from other components once the are produced. This contrasts the situation in CMB anisotropies, where photons and baryons are tightly coupled due to Compton scattering before the recombination; as a result, any pre-existing anisotropies are suppressed and information of initial state is erased. On the other hand, GW anisotropies can carry the memory of the initial condition due to the absence of the collision term. This unique feature may be employed to probe the shape of the primordial curvature perturbation~\cite{Li:2021iva,ding:2023xxx}.}
\item {The decoupling time of GW is considerably earlier than that of photon due to their distinct generating mechanisms, leading to different SW and ISW contributions in the GW and CMB cases. In addition, as demonstrated earlier, the SW contribution is also tightly related to the slope of the isotropic GW spectrum through the initial condition of monopole term. For example, the SW terms could be strongly suppressed when the slope $\gamma_{\rm{GW}}\sim 2$. This feature could help to distinguish various production mechanisms based on the behavior of the angular power spectrum at large scales.}
\end{enumerate}}
\item In contrast to the recombination time $\eta \sim 280~\rm{Mpc}$, the initial time $\eta_{\rm{in}}$ when GWs are produced and begin to freely propagate is significantly smaller, and it depends on the specific mechanism that generates GW. 
From Eqs.~\eqref{eq:29}\eqref{eq:30}, $\eta_{\rm{in}}$ influences the transfer functions through the metric perturbations and projection effects from Spherical Bessel functions, potentially introducing model dependence to the anisotropies in the CGWB.
However, in practice, we have observed that the varying specific value of $\eta_{\rm{in}}$ does not appreciably alter any anisotropic signals noticeably provided that $\eta_{\rm{in}}$ remains small (see also \cite{Schulze:2023ich}). We therefore adopt $\eta_{\rm{in}}=10^{-4}\,\mathrm{Mpc}$ in the rest part of this paper. 

\end{itemize}

\begin{figure}
\centering
\includegraphics[width=0.8\textwidth]{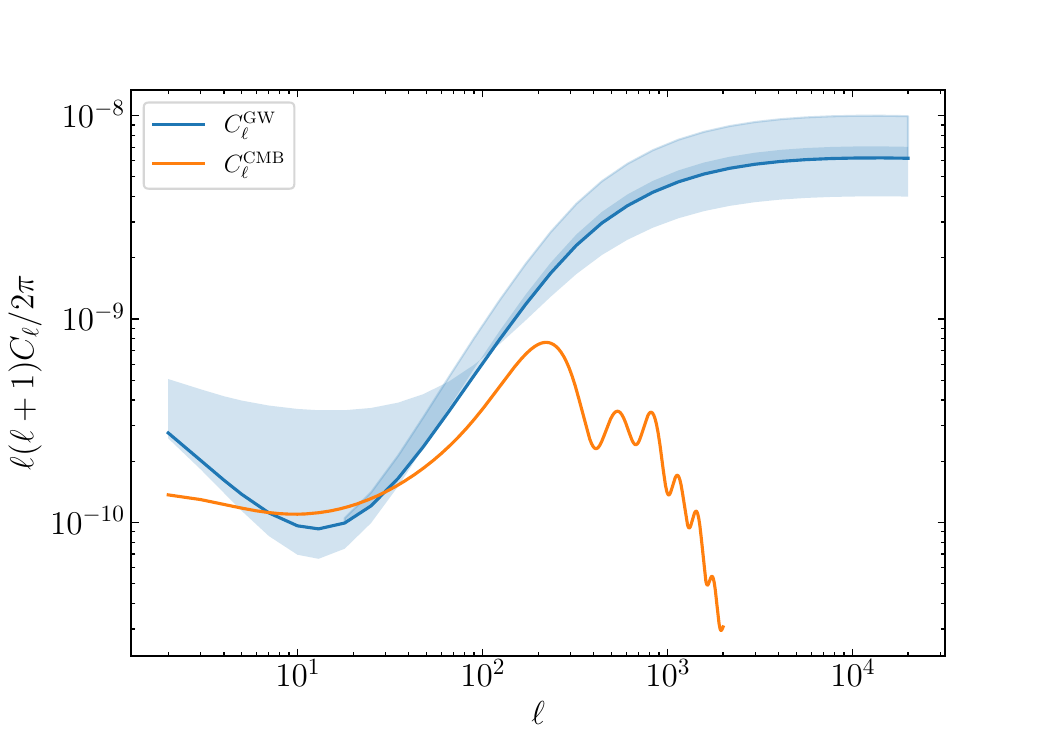}
\includegraphics[width=0.42\textwidth]{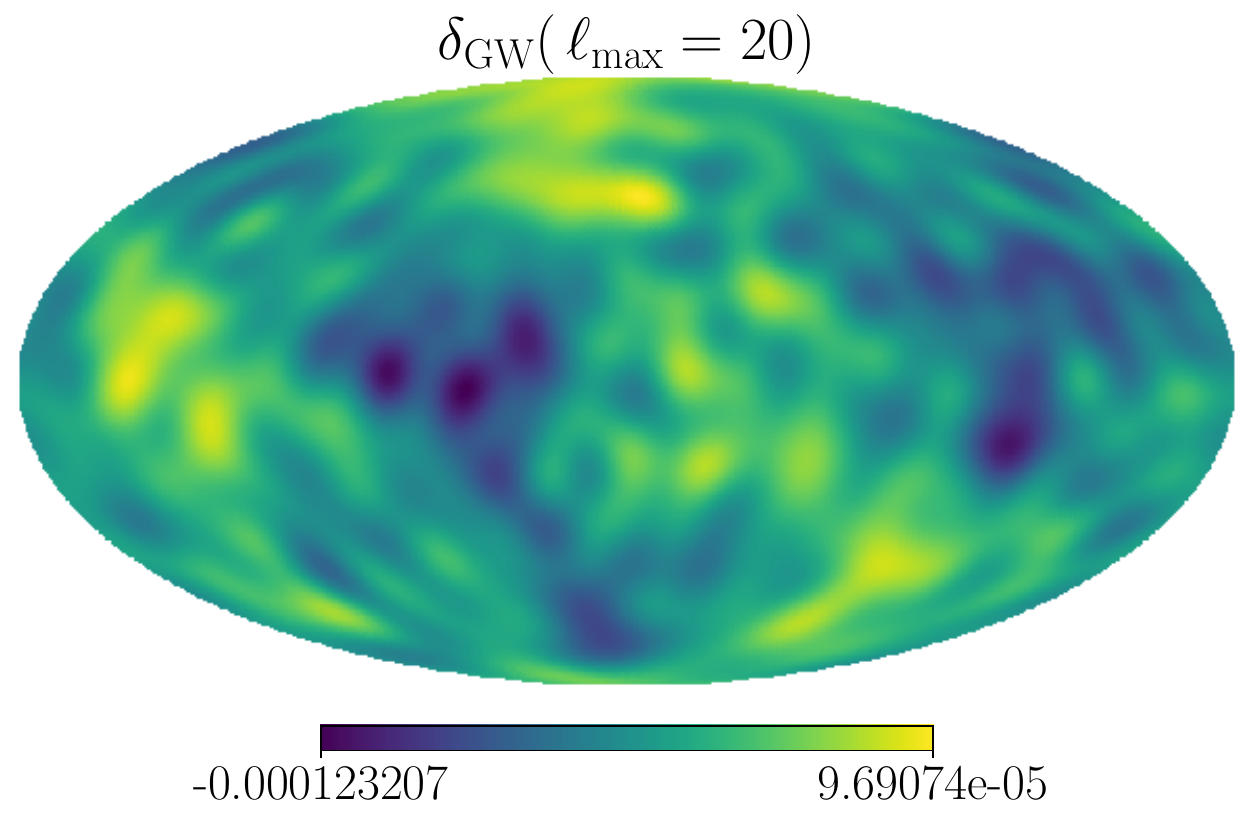}
\includegraphics[width=0.42\textwidth]{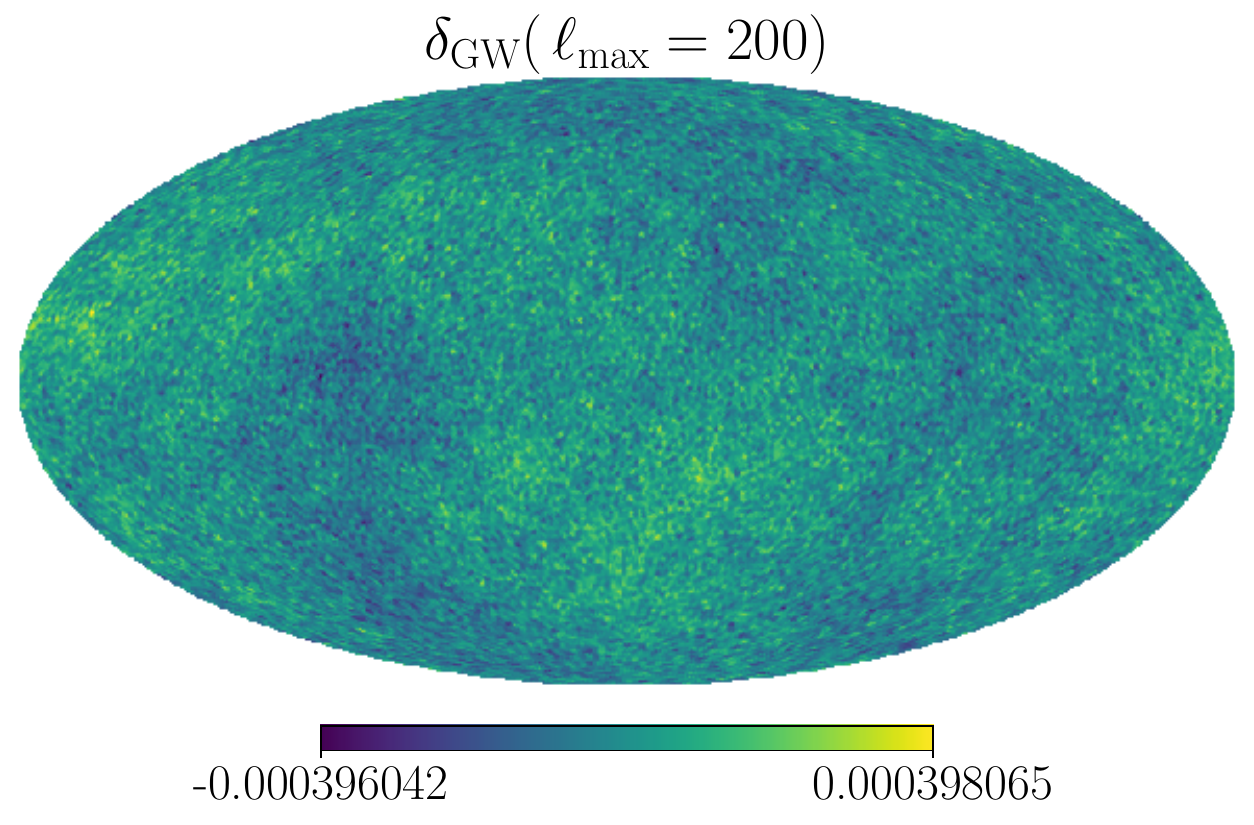}
\includegraphics[width=0.42\textwidth]{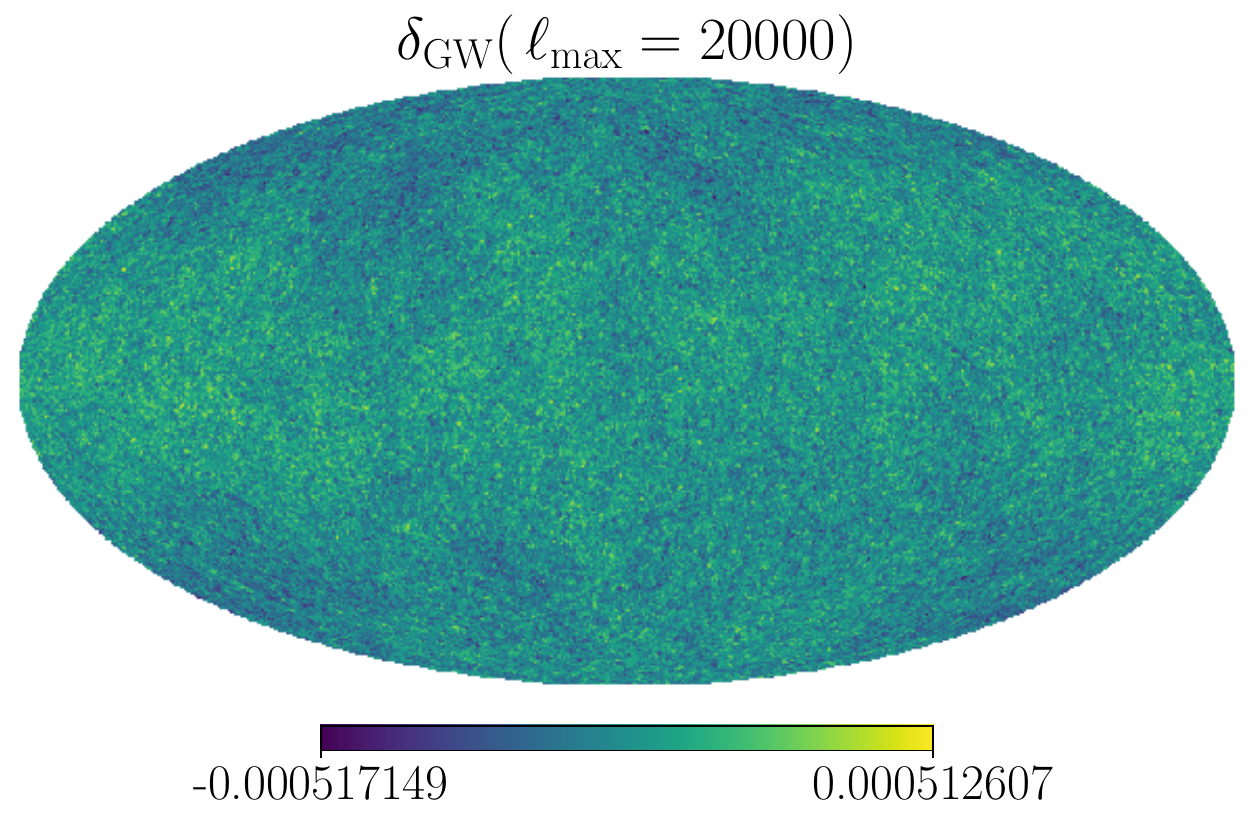}
\caption{\textbf{Upper panel}: The dimensionless CGWB angular power spectrum up to $\ell_{\rm{max}}=20000$ (blue solid line), which is computed from the best-fitted spectrum of the NANOGrav observation. The light blue band is computed based on $\gamma_{\rm{GW}}$ ranging within $68\%$ confidence interval. For comparison, the dimensionless CMB angular power spectrum (orange solid line) is also shown. \textbf{Lower panel}: HEALPix maps of $\delta_{\rm{GW}}$  with different cut-off multipoles $\ell_{\rm max}$.}.
\label{fig:GWCL}
\end{figure}

Based on the framework introduced above, we proceed to compute the transfer functions and the angular power spectrum of the CGWB. To perform the necessary numerical integrations, we rely on samplings of the metric perturbations obtained from the Boltzmann code~{\tt CLASS}~\cite{Lesgourgues:2011re,Blas:2011rf}, and the primordial power spectrum of scalar perturbations is employed with the amplitude and the spectral index are  $\ln \left(10^{10} A_s\right)=3.04$, $n_s=0.966$ respectively, with the standard value of a pivot scale $k_{\rm pivot}=0.05 \mathrm{Mpc}^{-1}$.
We present the dimensionless angular power spectrum of CGWB (Eq.~\eqref{eq:13}) up to $\ell_{\rm{max}}=20000$ in Fig.~\ref{fig:GWCL}. For comparison, the dimensionless CMB angular power spectrum is also shown, which is given directly by {\tt CLASS}. One can see that in analogous to the CMB, the CGWB is dominated by the monopole component, with small anisotropies of the order of $10^{-4}$. This is because the Boltzmann equations for gravitational waves (GWs) and CMB photons share an analogous form, except that the propagation of GWs has different initial conditions and lacks a collision term. As a result, the transfer functions of these two systems have a similar order of magnitude, which tends to restrict the size of GW fluctuations to be on the same order of magnitude as curvature perturbations. On the other hand, CGWB anisotropies have some special features different from CMB, which are summarized as follows:
\begin{itemize}
  \item { On large scales, the contribution of SW effect on CGWB anisotropies is highly suppressed, and the dominant contribution comes from ISW effect. It is well known that SW term dominates the CMB spectrum on large scale, where the observed CMB anisotropies scale as $\Theta_{\rm SW} \sim \delta_\gamma (\eta_*)/4+\Psi(\eta_*)=\Psi(\eta_*)/3$. However, as we have seen, the CGWB monopole component heavily depends on the slope of isotropic GW spectrum at $\eta_{\rm{in}}$, which gives its anisotropies a different behavior. To be specific, the favored value $\gamma_{\rm{GW}}\sim 1.82$ extracted from NANOGrav data implies a strong cancellation between intrinsic anisotropy and gravitational redshift, leading to very small SW anisotropies $\calG_{\rm SW} \sim 0.08 \Psi (\eta_{\rm{in}})$. As a consequence, in our case, CGWB angular power spectrum on large scales is dominated by early and late ISW effect, which result from the evolution of metric perturbations caused by residual amount of radiation at early times and due to dark energy becoming dominant at late times.}
  \item { On small scales, CGWB anisotropies increase smoothly due to gravitational effects. As we have observed, the CGWB anisotropies only depend on the time evolution of metric perturbations through the ISW term. Instead of being a constant during the matter-dominated era, the metric perturbation $\Psi$ decays on sub-horizon scales during the radiation-dominated era. This decay occurs rapidly, boosting the short-wavelength modes with $k>k_{\rm eq}\approx 0.01 \mathrm{Mpc}^{-1}$, corresponding to multipoles $\ell> k_{\rm eq}(\eta_0-\eta_{\rm{in}}) \approx 150$, as shown in Fig.~\ref{fig:GWCL}. In contrast, for the CMB scenario, photons and baryons remain tightly coupled due to the collision term until the last scattering surface, and acoustic oscillations arise on subhorizon scales due to the competition between radiation press and gravity, imprinting acoustic wiggles to the CMB spectrum. Additionally, anisotropies on scales smaller than the diffusion length$\lambda_D \sim (a\sqrt{n_e \sigma_T H})^{-1}$ are strongly damped due to the diffusion damping, where $n_e$ is the free electron number density and $\sigma_T$ the Compton scattering cross-section. This damping effect ultimately erases the gravitational boost in high $\ell s$ of CMB spectrum.}
\end{itemize}

\section{Cross-correlations of CGWB with CMB and gravitational lensing}
\label{sec:cross}

 In this section, we further study the cross-correlations of CGWB anisotropies with CMB tracers, which serve as equally important observables. We focus on two specific correlations: CMB temperature anisotropies and CMB lensing. Since the GW and the CMB photon share the same geodesics under the same metric perturbation during their propagation, we expect that CGWB anisotropies have a noticeable correlation with these two CMB tracers. The cross-correlation between CGWB and CMB has been previously studied in~\cite{Ricciardone:2021kel,Braglia:2021fxn,Schulze:2023ich}, we here give the result based on NANOGrav data and show that it significantly affects the low $\ell$ of CGWB$\times$CMB angular power spectrum. The $a_{\ell m}$s of CMB temperature fluctuations are 
\begin{align}
a^{\rm CMB}_{\ell m}(\vec{x}, \eta) &= \int d\Omega_{\hat{n}} Y_{\ell m}^*(\hat{n})\Theta_{\ell}(\vec{x},\hat{n},\eta) \nonumber\\
&= \sum_{\ell'} (-i)^{\ell'} (2\ell' + 1)\int\frac{d^3k}{(2\pi)^3}e^{i\vec{k}\cdot\vec{x}}\int d\Omega_{\hat{n}} Y_{\ell m}^*(\hat{n})\mathcal{P}_{\ell'}(\mu)\Theta_{\ell'}(\vec{k},\eta),
\end{align}
and the line-of-sight solution of CMB multipoles reads~\cite{Dodelson:2003ft}
\begin{align}
\Theta_{\ell}(k, \eta_*, \eta_0) &= \int_{\eta_*}^{\eta_0} d \eta \left\{g(\eta)\left[\Theta_0(k, \eta)+\Psi(k, \eta)\right] j_{\ell}\left[k\left(\eta_0-\eta\right)\right] +g(\eta) \frac{-i v_b(k, \eta)}{k}  j_{\ell}^{\prime}\left[k\left(\eta_0-\eta\right)\right]\right.   \nonumber\\
& \left.+e^{-\tau(\eta)}\left[\Psi^{\prime}(k, \eta) + \Phi^{\prime}(k, \eta)\right] j_{\ell}\left[k\left(\eta_0-\eta\right)\right]\right\} \nonumber\\
&\simeq \underbrace{\left[\Theta_0(k, \eta_*)+\Phi(k, \eta_*)\right]}_{\rm{SW}} j_{\ell}\left[k\left(\eta_0-\eta_*\right)\right] + \underbrace{\frac{-i v_b(k, \eta_*)}{k}}_{\rm{Doppler}} j_{\ell}^{\prime}\left[k\left(\eta_0-\eta_*\right)\right]   \nonumber\\
& + \underbrace{\int_{\eta_*}^{\eta_0} d \eta \left[\Psi^{\prime}(k, \eta) + \Psi^{\prime}(k, \eta)\right]}_{\rm{ISW}} j_{\ell}\left[k\left(\eta_0-\eta\right)\right],
\label{eq:33}
\end{align}
where $\tau(\eta) = \int_\eta^{\eta_0} d \eta\,n_e \sigma_T a$ is the optical depth, $v_b$ is the velocity of baryons, and $g(\eta) = -\tau^{\prime}(\eta) e^{-\tau(\eta)}$ is the visibility function. We have also used the instantaneous recombination assumptions $g(\eta)= \delta(\eta-\eta_*)$ and $e^{-\tau}=\theta(\eta-\eta_*)$ to obtain a simpler expression in the last line of Eq.~\eqref{eq:33}, where $\theta$ represents the Heaviside function. Eq.~(\ref{eq:33}) indicates that CMB anisotropies are characterized by the sum of the SW, Doppler and ISW contributions. Based on this analysis, the corresponding transfer functions can be extracted as follows:
\begin{align}
\Delta^{\rm CMB}_\ell(k, \eta_*, \eta_0) &= \Delta_{\ell}^{\mathrm{SW}}\left(k, \eta_*, \eta_0\right) + \Delta_{\ell}^{\mathrm{DOP}}\left(k, \eta_*, \eta_0\right) + \Delta_{\ell}^{\mathrm{ISW}}\left(k, \eta_*, \eta_0\right), \label{eq:34}\\
\Delta_{\ell}^{\mathrm{SW}}\left(k, \eta_*, \eta_0\right) &= \frac{\left[\Theta_0(k, \eta_*)+\Phi(k, \eta_*)\right]}{\mathcal{R}(\vec{k})} j_{\ell}\left[k\left(\eta_0-\eta_*\right)\right], \label{eq:35}\\
\Delta_{\ell}^{\mathrm{DOP}}\left(k, \eta_*, \eta_0\right) &= -\frac{i v_b(k, \eta_*)}{k \mathcal{R}(\vec{k})} j_{\ell}^{\prime}\left[k\left(\eta_0-\eta_*\right)\right] \nonumber\\
&= 3 \Theta_1\left(k, \eta_*\right)\left(j_{\ell-1}\left[k\left(\eta_0-\eta_*\right)\right]-(\ell+1) \frac{j_\ell\left[k\left(\eta_0-\eta_*\right)\right]}{k\left(\eta_0-\eta_*\right)}\right), \label{eq:36}\\
\Delta_{\ell}^{\mathrm{ISW}}\left(k, \eta_*, \eta_0\right) &= \int_{\eta_*}^{\eta_0} d \eta \frac{\left[\Psi^{\prime}(k, \eta) + \Phi^{\prime}(k, \eta)\right]}{\mathcal{R}(\vec{k})} j_{\ell}\left[k\left(\eta_0-\eta\right)\right].  \label{eq:37}
\end{align}
Inserting above equations and GW transfer function~(\ref{eq:28})-(\ref{eq:30}) in the definition of cross-correlation angular spectrum $\langle a^{\rm{GW}}_{\ell m} a^{\rm{CMB}*}_{\ell^{\prime} m^{\prime}}\rangle \equiv \delta_{\ell \ell^{\prime}} \delta_{m m^{\prime}} C_{\ell}^{\rm{GW}\times \mathrm{CMB} }$ yields the cross-correlation spectrum in  the form of summation of six terms:
\begin{align}
C_{\ell}^{\mathrm{GW} \times \mathrm{CMB}} &= 4\pi\int\frac{dk}{k}\mathcal{P}_\mathcal{R}(k) \left[ T^{\rm GW}_\ell(k, \eta_{\rm{in}}, \eta_0)\,\Delta^{\rm CMB}_\ell(k, \eta_*, \eta_0) \right] \nonumber\\
&=C_{\ell}^{\mathrm{SW} \times \mathrm{SW}}+C_{\ell}^{\mathrm{SW} \times \mathrm{DOP}}+C_{\ell}^{\mathrm{SW} \times \mathrm{ISW}}+C_{\ell}^{\mathrm{ISW} \times \mathrm{SW}}++C_{\ell}^{\mathrm{ISW} \times \mathrm{DOP}}+C_{\ell}^{\mathrm{ISW} \times \mathrm{ISW}}.
\end{align}

\begin{figure}
\centering
\includegraphics[width=0.8\textwidth]{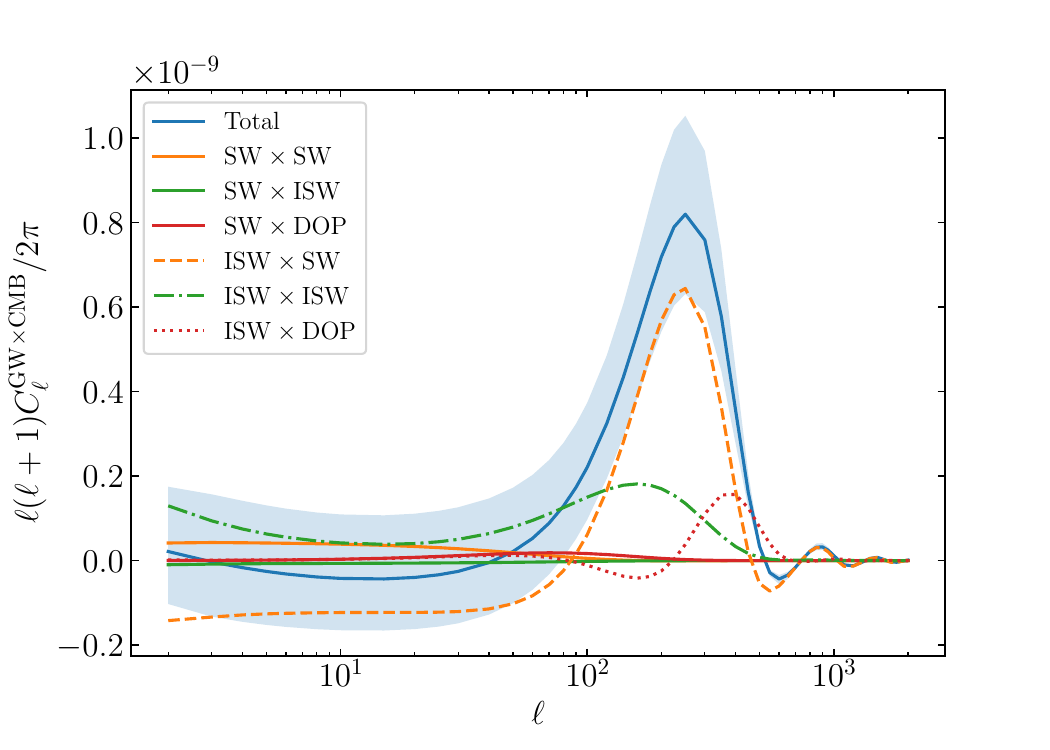}
\caption{Various contributions to the CGWB $\times$ CMB cross-correlation
angular power spectrum. The light blue band is computed based on $\gamma_{\rm{GW}}$ ranging within $68\%$ confidence interval.}
\label{fig:cmb}
\end{figure}

For completeness, we plot each cross-spectrum based on the central value of $\gamma_{\rm GW}$ from the NANOGrav data in Fig.~\ref{fig:cmb}. Firstly, we observe that all correlations associated with the SW terms of CGWB are substantially suppressed. This is anticipated, considering the SW effects in the CGWB auto-spectrum are also sub-dominant\footnote{Note that if the SW term in CGWB was not suppressed, in contrast to the CMB, the SW $\times$ DOP contribution in the CGWB $\times$ CMB signal should be noticeable. This is because the distance $\eta_{\rm{in}}-\eta_*$ shifts the SW term of the CGWB, which restores the coherence between the SW and the Doppler terms.}. As a result, the total cross-correlations at low $\ell$s ($\ell < 40$) tend to be marginally negative, due to the cancellation of ISW$\times$ISW and ISW$\times$SW terms. This lack of cross-correlation between CGWB and CMB signal is one of the implications given by PTA observations. At high $\ell$s, the total cross-spectrum is dominated by the ISW cross-correlations of the CGWB and behaves similarly to the CMB case. For ISW$\times$SW term, it has a peak around $l\sim200$ which coincides with the first acoustic oscillation and then damps at smaller scales. For ISW$\times$ISW term, it forms a tilted plateau at large scales resulting from the late ISW effect, and has a peak near the first acoustic peak due to the early ISW effect.

The anisotropic CGWB can also be cross-correlated with the CMB lensing signal, which is generated from the gravitational deflection of CMB photons due to large-scale structure. At the linear level, the lensing potential can be defined as
\begin{align}
\psi = -2 \int_{\eta_0 - \eta_*}^{\eta_0} \mathrm{d} \eta \frac{(\eta - \eta_{*})}{(\eta_0 - \eta_{*})(\eta_0 - \eta)} \Psi(k, \eta),
\end{align}
so that the lensing convergence can be computed through 
\begin{align}
\label{eq:kappa}
    \kappa = -\frac12 \nabla^{(2)}\psi.
\end{align}
After multiplying by a radial length, the 2-dimensional Laplacian in Eq.~\eqref{eq:kappa} can be replaced by the 3-dimensional one:  by ignoring radial modes, which introduce a $k^2$ term in the transfer function. 

Therefore, we can easily write down the auto- and cross-correlations spectrum between the CGWB and the CMB lensing potential and lensing convergence as:
\begin{align}
C_l^{\kappa \kappa} &= 4 \pi \int \mathrm{d} k k^3 \mathcal{P}_{\mathcal{R}}(k) \Big|\Delta^{\kappa}_\ell(k, \eta_{*}, \eta_0) \Big|^2, \\
C_l^{\mathrm{GW} \times \kappa} &= 4 \pi \int \mathrm{d} k k \mathcal{P}_{\mathcal{R}}(k)\left[ T^{\rm GW}_\ell(k, \eta_{\rm{in}}, \eta_0)\,\Delta^{\kappa}_\ell(k, \eta_*, \eta_0)\right],
\end{align}
where the corresponding transfer functions are
\begin{align}
\Delta^{\kappa}_\ell(k, \eta_*, \eta_0) &= \int_{\eta_0 - \eta_*}^{\eta_0} \mathrm{d} \eta \frac{\Psi(k, \eta)}{\mathcal{R}(\vec{k})} \frac{(\eta_0 - \eta)(\eta - \eta_{*})}{\eta_0 - \eta_{*}}\,j_{\ell}\left[k\left(\eta_0-\eta\right)\right].
\end{align}

\begin{figure}
\centering
\includegraphics[width=0.8\textwidth]{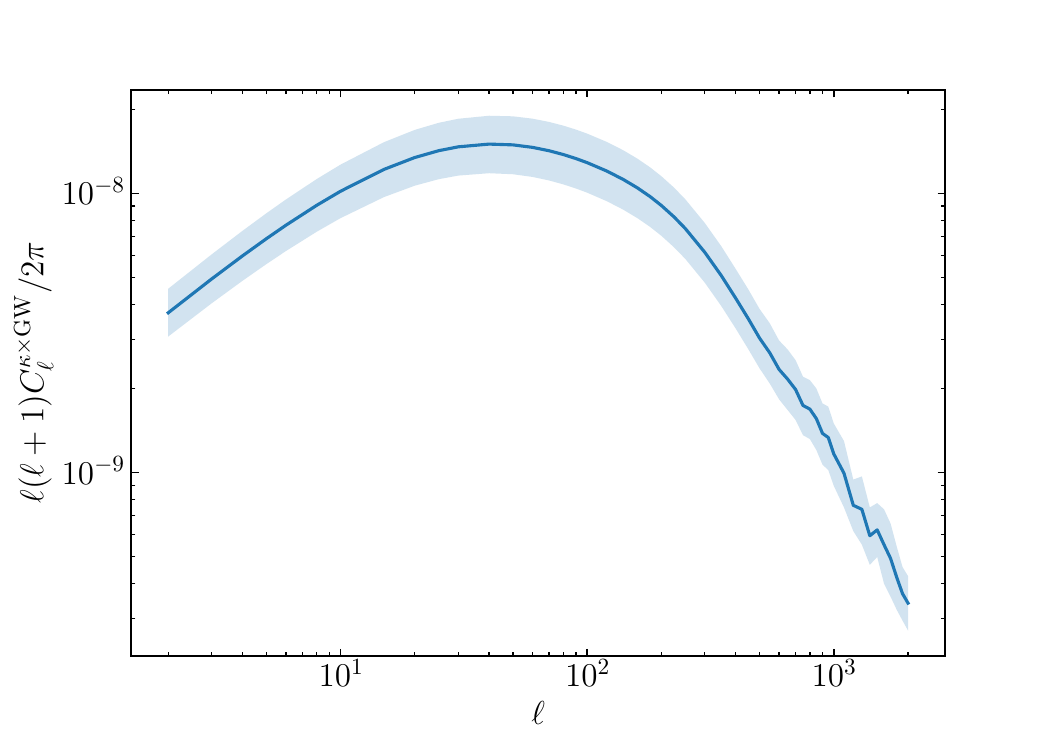}
\includegraphics[width=0.8\textwidth]{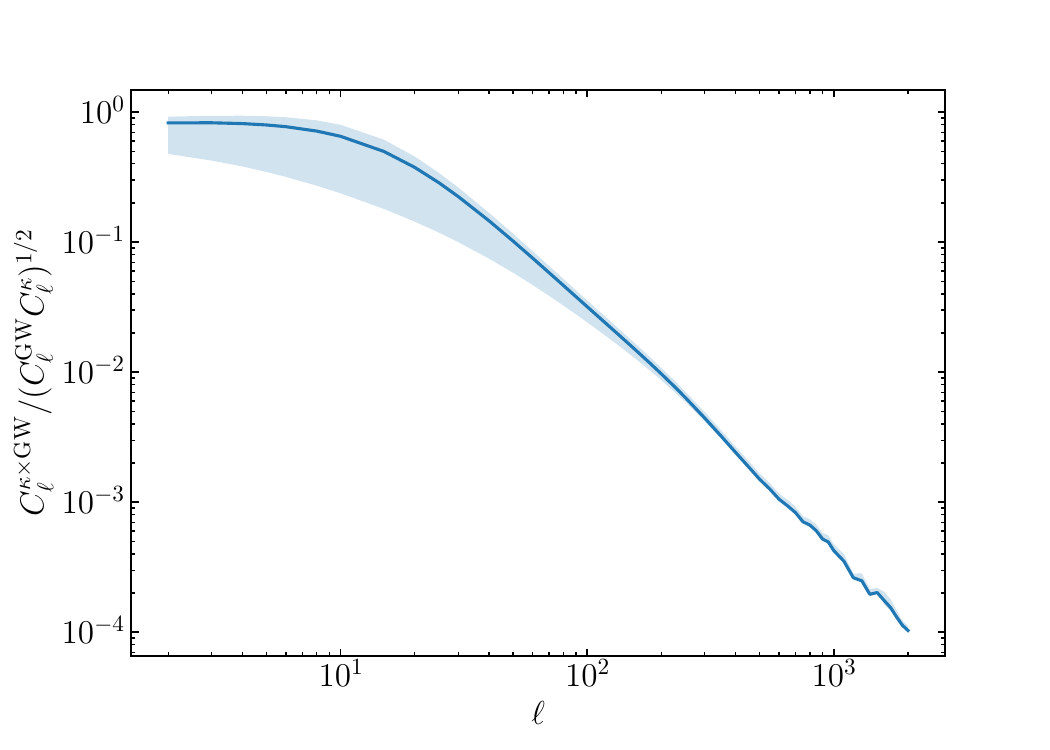}
\caption{\textbf{Upper panel}: Cross-correlation spectrum of the lensing convergence with the CGWB. \textbf{Lower panel}: Normalized cross-correlation spectrum. The light blue band is computed based on $\gamma_{\rm{GW}}$ ranging within $68\%$ confidence interval.}
\label{fig:lensing}
\end{figure}

We compute the cross-correlation of the CGWB with the CMB lensing and present results in Fig.~\ref{fig:lensing}. The upper panel shows that the cross-correlation peaks at around $\ell\sim60$, which coincides with the scale where the spectrum of lensing potential peaks (see \cite{Lewis:2006fu}), indicating a direct connection. The normalized cross-correlation (lower panel in Fig.~\ref{fig:lensing}) shows significant correlations at large scales, which however quickly drops to less than $10\%$ at $\ell\sim60$, where the CMB lensing convergence peaks. However, by comparing the upper panel in Fig.~\ref{fig:lensing} to Fig.~\ref{fig:cmb}, we notice that despite the low normalized correlation, the peak strength of the absolute correlations with the lensing convergence is still an order of magnitude higher than the cross-correlation with the CMB temperature. This suggests that compared with the CGWB $\times$ CMB, the CGWB $\times$ lensing single could potentially be a better tracer to identify possible anisotropic signals in the CGWB. 

\section{Discussion and conclusion}
\label{sec:diss}
The CGWB now plays an increasingly important role in cosmology because its generating mechanisms are closely related to physics beyond the standard model, and it directly carries information of early Universe. In contrast to the CMB, the CGWB may carry unique anisotropic signals, which originate from the fact that the CGWB is generally produced much earlier than the CMB and remains collisionless during propagation, making it a powerful tool to probe various aspects of early universe physics. 

In this work, we focus on the latest SGWB results reported by PTA  observations and discuss their implications on the SGWB anisotropies in a model-independent way. We compute the angular power spectrum of anisotropies in the CGWB implied by the recent NANOGrav 15-year data to high multipoles, where unsupressed information of curvature perturbations is present. Additionally, we investigate in detail the correlations between the CGWB and CMB anisotropies, as well as the CMB lensing. Notably, our findings suggest that the suppressed SW effects, as implied by recent NANOGrav data, lead to a tilted plateau in the CGWB auto-correlation spectrum, and a marginally negative signal in the CGWB$\times$CMB cross-correlation spectrum at large scales. These distinctive patterns in the CGWB anisotropies represent one of the important implications of the NANOGrav data which has not been reported by previous study.  On the other hand, we also compute the cross-correlation between the anisotropies in the CGWB and the CMB lensing data. We pointed out that the high cross-correlations between these two tracers are maintained at lower multipoles based on the recent NANOGrav data. This suggests that, similar to the strategies in \cite{Ricciardone:2021kel}, some reconstruction techniques for the CGWB anisotropic maps could be performed based on the CMB lensing data, which will benefit any future detection and foreground cleaning of the anisotrpic GWB signals.

As we have pointed out, the most unique features of the CGWB, including a tilted plateau on low $\ell$s due to the suppressed SW effect and the enhancement on high $\ell$s due to the absence of the diffusion damping, concentrate on the scales $\ell_{\max }\sim 20$ and $\ell_{\max }>1000$. 
However, even though we have made historical progress in detecting the  SGWB, it remains challenging to detect these anisotropies~\cite{NANOGrav:2023tcn}. 
This is because, for any PTA observations based on $N_{\rm p}$ pulsars, the total number of time residual correlations extracted from the pulsars, in the form of overlap reduction function, is given by $N_{\rm{cc}}=N_{\rm p}\left(N_{\rm{p}}-1\right)/2$.  To extract the anisotropic components, we can expand the SGWB in the form of overlap reduction function with spherical-harmonic function, analogous to the approach used to study the CMB. Using this method, one can roughly estimate the required pulsar numbers. For a maximum resolvable multipole $\ell_{\max }$, the corresponding degrees of freedom of the spherical-harmonic function are given by $\sum_{\ell=0}^{\ell_{\max }}(2 \ell+1)=\left(\ell_{\max }+1\right)^2 \sim N_{\rm{cc}}$, which implies that $N_{\rm{p}} \sim \ell_{\max }$. The future PTA observations such as SKA are expected to have a significant improvement, potentially observing the CGWB spectrum at $\ell_{\max }< 20$. Although the detection may be still challenging due to the decreasing sensitivity at higher $\ell$s. 

On the other hand, for the angular scales with $\ell_{\max }>1000$, the challenge becomes even more significant. 
This difficulty is currently beyond the capability of any current and future PTA experiments. This challenge motivates us to explore other observations. As pointed out in Refs.~~\cite{Moore:2017ity,Klioner:2017asb,Mihaylov:2019lft,Jaraba:2023djs}, it is also possible to detect the SGWB using astrometric measurements. Analogous to the passage of GWs propagating over the Earth-pulsar system, affecting the pulsar timing residuals, the apparent position of a star also changes when GWs pass over the Earth-star system. This potentially allows us to extract the monopole spectrum and anisotropies of the SGWB from astrometric data sets. The European Space Agency (ESA) mission Gaia~\cite{Gaia:2016zol} is now providing an all-sky astrometric map of $N_{\mathrm{star}} \sim 10^9$ stars, which in principle, have capability to detect more small scale CGWB anisotrpies. 

In addition, our methodology does not account for distinct imprints that might be associated with specific sources. For example, as highlighted in~\cite{Geller:2018mwu,Kumar:2021ffi,Bodas:2022urf,ValbusaDallArmi:2023nqn}, certain sources of the CGWB, such as the phase transitions in the presence of isocurvature perturbations or the metric perturbations resulting from quantum fluctuations during inflation, may cause  the standard adiabatic initial condition in Eq.~\ref{eq:27} not to hold and introduce additional anisotropies to the CGWB. These results therefore suggest careful examinations of various CGWB generation mechanisms are necessary. We reserve detailed investigations on this topic in our future study.

\noindent {\bf Acknowledgments.~}
We would like to thank Shuailiang Ge, Huai-Ke Guo, Fapeng Huang, Yongping Li, Shi Pi, and Junchao Zong for useful discussions. We also acknowledge the use of the ${\tt NumPy}$ and ${\tt SciPy}$ scientific computing packages. Ran Ding is supported in part by the National Key R\&D Program of China (2021YFC2203100). Chi Tian is supported by the Natural Science Foundation of Anhui Province (Grants No. 2308085QA34). The authors acknowledge the High-performance Computing Platform of Anhui University for providing computing resources. 
\bibliography{Ref}

\end{document}